\documentclass{PoS}

\usepackage{color}
\usepackage{colordvi}

\usepackage[normalem]{ulem}

\usepackage{color}
\newcommand{\beqn}{\begin{eqnarray}}
\newcommand{\eeqn}{\end{eqnarray}}
\newcommand{\be}{\begin{equation}}
\newcommand{\ee}{\end{equation}}

\newcommand{\ba}{\begin{array}{c}}
\newcommand{\bat}{\begin{array}{cc}}
\newcommand{\ea}{\end{array}}

\newcommand{\bi}{\begin{itemize}}
\newcommand{\ei}{\end{itemize}}

\newcommand{\ket}{\,\rangle}
\newcommand{\bra}{\langle \,}

\newcommand{\Frac}[2]{\frac{\displaystyle #1}{\displaystyle #2}}

\newcommand{\cO}{{\cal O}}

\newcommand{\mF}{\mathcal{F}}
\newcommand{\mG}{\mathcal{G}}
\newcommand{\mH}{\mathcal{H}}

\newcommand{\mL}{\mathcal{L}}
\newcommand{\mM}{\mathcal{M}}

\newcommand{\gsim}{\stackrel{>}{_\sim}}

\newcommand{\bear}{\begin{eqnarray}}
\newcommand{\eear}{\end{eqnarray}}
\newcommand{\nn}{\nonumber}

\title{
\vspace*{-6.cm}\hspace*{8cm}
{\small 
 FTUAM-15-42,
 IFT-UAM/CSIC-15-121
}
\vspace*{4.cm}\hspace*{-15cm}
\\
Composite resonances and their impact on the electroweak chiral Lagrangian
}

\ShortTitle{Composite resonances and their impact on the EW chiral Lagrangian}

\author{\speaker{Juan Jos\'e Sanz-Cillero}
\thanks{
I would like to thank the organizers for the nice scientific and social
environment during the Conference.
Lots of thanks also to A. Pich, I. Rosell and J. Santos for their help
in the preparation of these proceedings.
This work has been supported by ERDF funds from the European Commission
[FPA2013-44773-P, SEV-2012-0249, CSD2007-00042].
}\\
        Departamento de F\'isica Te\'orica and Instituto de F\'isica Te\'orica,
IFT-UAM/CSIC,\\
Universidad Aut\'onoma de Madrid, Cantoblanco, 28049 Madrid, Spain
\\
        E-mail: \email{juanj.sanz@uam.es}}

\abstract{
In this talk we study the low-energy effective couplings generated by
strongly-coupled electroweak models that contain heavy composite resonances.
Invariance under $SU(2)_L\times SU(2)_R$ is a key ingredient in the construction of the
resonance  action.
For simplicity, in these proceedings we focus our attention on the impact of
a heavy colourless vector $V$, which transforms as
a triplet under the custodial group. More precisely, we study the couplings that are relevant
for the vector form-factors of the $L+R$ current into two electroweak Goldstones
and into two Standard Model fermions, which contribute to
the oblique parameters $S$ and $T$ and the anomalous
$Z\to f\bar{f}$ couplings, respectively. Our predictions are compatible
with bounds from direct and indirect searches for $M_V\gsim 1.5$~TeV.
Finally, although we consider an antisymmetric tensor formalism to describe the vector resonance,
we derive the equivalent action in the Proca four-vector representation
and show that the predictions for low-energy couplings and form-factors are identical, as expected.
}

\FullConference{The European Physical Society Conference on High Energy Physics\\
		22--29 July 2015\\
		Vienna, Austria}

\begin{document}

\section{Impact of heavy resonances on the low-energy electroweak effective theory}

So far the Large Hadron Collider (LHC) has not found any trace of beyond
the Standard Model (BSM) states with masses below 1~TeV.
Likewise, no significant deviation
has been observed in the low-energy interactions between Standard Model (SM) particles.
Effective field theories are then the natural approach.
In this talk~\cite{Santos:2015,Pich-preparation} we discuss
the possibility of strongly-coupled BSM scenarios
with the approximate custodial symmetry invariance
of the SM, exact in the SM scalar sector.
We develop an
invariant Lagrangian under $\mG=SU(2)_L\times SU(2)_R$, which spontaneously breaks down to
the custodial subgroup $\mH=SU(2)_{L+R}$ and generates
the electroweak (EW) would-be Goldstone bosons $\varphi^a$,
described a unitary $2\times 2$ matrix $U(\varphi)$.
In these (non-linear) EW chiral Lagrangian with a light Higgs (ECLh), the low-energy amplitude $\mM$ has an expansion in powers of infrared scales $p$
(external momenta and SM masses)
of the form
(e.g., for $2\to 2$ processes)~\cite{Pich-preparation,Weinberg:1978kz,Georgi-Manohar,Buchalla:2013eza,Guo:2015},
\bear
\mM &\sim & \underbrace{ \Frac{p^2}{v^2}  }_{\mbox{LO (tree)}}
\, + \, \bigg(
\, \underbrace{ a_{k}^r }_{\mbox{ NLO (tree) }} \quad -\quad
\underbrace{   \Frac{ \Gamma_{k} }{16\pi^2}\ln\Frac{p}{\mu} \quad +\quad ...  }_{
\mbox{NLO (1-loop)}    }
\quad
\bigg)
\,\,\, \Frac{p^4}{v^4}\, \,\, +\,\,\, \cO(p^6) \, .
\label{eq.chiral-amp}
\eear
The EW effective theory (EWET) Lagrangian operators can be sorted out based on their chiral dimension:
\bear
\mL_{\rm EWET}\,=\, \mL_2\, +\, \mL_4\, +\, ...
\eear
where the operators in $\mL_{\hat{d}}$ are of
$\cO(p^{\hat{d}})$~\cite{Pich-preparation,Weinberg:1978kz,Georgi-Manohar,Buchalla:2013eza}.
Covariant derivatives and masses are $\cO(p)$~\cite{chpt}
and each fermion field scales like $\cO(p^{1/2})$ in naive dimensional
analysis (NDA)~\cite{Pich-preparation,Georgi-Manohar,Buchalla:2013eza,Buchalla:2013rka}.
The $\mG$--invariant operators in $\mL_{\rm EWET}$ are built with the Goldstone tensors $U(\varphi)$,
functions $\mF_k$ of the Higgs singlet $h$, its derivatives $\partial_{\mu_1}...\partial_{\mu_n} h$,
the gauge fields and the SM fermions $\psi$~\cite{Buchalla:2013rka,Longhitano:1980iz,Morales:94,SILH,Alonso:2012,Grinstein:2007iv}.
{}From the chiral counting point of view $\mL^{\rm SM}$ would be $\cO(p^2)$ but its underlying
renormalizable structure makes all $\Gamma_k=0$ and ensures the absences of
higher-dimension divergences~\cite{Guo:2015,Alonso:2015}.
The most important contributions to a given process
are given by the operators of lowest chiral dimension.
The leading order (LO) contribution is $\cO(p^2)$ and is given by tree-level diagrams
with only $\mL_2$ vertices.
Likewise, the one-loop contribution with only $\mL_2$ vertices is $\cO(p^4)$;
it is suppressed in~(\ref{eq.chiral-amp}) with respect to
the LO by a factor $p^2/\Lambda_{\rm NL}^2$,
with $\Lambda_{\rm NL}^2\sim 16\pi^2 v^2 \Gamma_k^{-1}\gsim 3$~TeV
(with $v=(\sqrt{2} G_F)^{-1/2}\approx 246$~GeV).
This suppression factor is related to the non-linearity of the ECLh    %%%EWET
and $\Lambda_{\rm NL}\to \infty$ when the Higgs can be embedded in a complex doublet $\Phi$~\cite{Guo:2015}.~\footnote{
Ref.~\cite{Alonso:2015} provides a geometrical interpretation in terms of the curvature of
metric of the internal weak space of the Higgs. In the flat-space limit
one has $\Lambda_{\rm NL}\to \infty$. Linear-Higgs scenarios with a complex Higgs doublet $\Phi$
correspond to this case. True ``non-linear models'' are defined by a non-zero curvature,
not by their (non-linear) representation.
}

In these proceedings~\cite{Santos:2015,Pich-preparation} we focus our attention on the tree-level
next-to-leading order (NLO) contributions. They are $\cO(p^4)$ and are
provided by tree-level diagrams with
one $\mL_4$ vertex with low-energy coupling $a_k$ (LEC) and an arbitrary number of $\mL_2$ vertices.
%
%The $\cO(p^4)$ LECs encode the underlying physics at  short distances.
%In particular,
They get contributions from tree-level heavy resonance exchanges.
At low energies, these $\cO(p^4)$ terms in~(\ref{eq.chiral-amp}) are
typically suppressed with respect to the LO amplitude, $\cO(p^2)$,
by a factor $a_kp^2/v^2 \sim p^2/M_R^2$~\cite{Santos:2015,Pich-preparation,Ecker:1988te,Ecker:1989yg}.
%The renormalized low-energy constants (LECs)
%also get one-loop contributions from the underlying theory with resonances, which may be computed following
%a similar procedure to that employed in QCD~\cite{one-loop-rcht}.
%

At high energies, one must include both the light dof (SM particles)
and the possible composite resonances as active degrees of freedom (dof)
in the Lagrangian~\cite{Santos:2015,Pich-preparation,pseudovector-Cata}:
%
%we study
%colourless vector, axial-vector, scalar and pseudoscalar resonances.
%We consider custodial singlet ($R=V,A,S,P$)  and triplet  ($R_1=V_1,A_1,S_1,P_1$) multiplets.
%Exotic resonances with $J^{PC}=1^{+-}$ quantum numbers
%were studied in~\cite{pseudovector-Cata}.
%This high-energy Lagrangian has the generic form
%
\bear
\mL %%%_{\rm high}
&=& \mL_{\rm non-res}  %%%\, +\, \mL_R^{\rm Kin}
\, +\, \mL_R\, ,
\eear
where
$\mL_{\rm non-res}$ contains only SM fields and
$\mL_R$ is the part of the Lagrangian that also contains resonances~\cite{Santos:2015}.
%%%
%%%In the purely bosonic case $\mL_{\rm non-res}=\mL_2$~\cite{Santos:2015,Ecker:1989yg}.
%%% However, additional $\cO(p^4)$
%%%fermionic operators are needed in the full case~\cite{Pich-preparation}.
%%%
The part of the interaction Lagrangian $\mL_R$ relevant for our analysis
of the $\mL_4$ LECs is given by the terms linear in the resonance
fields,
$\Delta \mL_R =\, \,  R \,\,\mathbb{O}_{p^2}[\chi,\psi]$~\cite{Santos:2015,Pich-preparation,Ecker:1988te,Ecker:1989yg,pseudovector-Cata},
%\bear
%\Delta \mL_R &=& \, \,  R \,\, \cO_{p^2}[\chi,\psi]\, ,
%\eear
with $\chi$ ($\psi$) referring to the light bosonic (fermionic) fields.
The tensor $\mathbb{O}_{p^2}[\chi,\psi]$ that couples the heavy resonance $R$ to the
light dof is going to provide the first correction to the low-energy ECLh  %%%EWET
by means
of diagrams where one has a heavy resonance propagator $\sim 1/M_R^2$ exchanged between
two vertices with $\mathbb{O}_{p^2}[\chi,\psi]$.
This gives an EWET operator of $\cO(p^4)$.
%%%, avoiding
%%%large corrections to the SM Lagrangian $\mL^{\rm SM} \subset
%%%\mL_{2}$.
At low energies, resonance operators with tensors $\mathbb{O}[\chi,\psi]$ of a higher order in $p$
or containing two or more $R$ fields contribute only to $\mL_{\hat{d}}$ with $\hat{d}>4$.
%
%The same applies to resonance operators with two or more resonance fields: they can only appear
%in tree-level diagrams where there are two or more resonance exchanges, leading to
%a stronger low-energy suppression by $\sim 1/(M_R^2)^{n\geq 2}$.

%The formal procedure to integrate out the heavy $R$ fields in the generating functional~\cite{Ecker:1988te}:
%\bear
%e^{i S[\chi,\psi]^{\rm EWET}} &=& \,\Int [dR]\,\, e^{i S[\chi,\psi, R] }
%\quad\stackrel{\rm tree-level}{=}\quad e^{i S[\chi,\psi,R_{\rm cl}(\chi,\psi)]}\, .
%\eear
%
The tree-level contribution to $\mL_{\rm EWET}[\chi,\psi]$
is given by the underlying high-energy action $S[\chi,\psi, R]$
with the resonance fields $R$ evaluated at the classical solution $R_{\rm cl}(\chi,\psi)$
of their equations of motion (EoM).
Solving the resonance EoM and expanding their solutions in powers of momenta for $p\ll M_R$,
one can write the heavy fields as local operators of the EWET dof~\cite{Ecker:1988te}.
This prediction for the contribution to the low-energy ECLh  %%%EWET
can be complemented through
the consideration of
ultraviolet-completion hypotheses (sum-rules~\cite{WSR,Peskin:92}, unitarity~\cite{Ecker:1989yg},
asymptotic form-factor counting rules~\cite{Brodsky-Lepage}...).
This imposes constraints on the resonance couplings that then turn into predictions
for the low-energy theory.

\section{Phenomenological example: vector form-factors}

Let us illustrate this with a basic example. We consider a colourless triplet vector resonance $V$
in a composite theory with the same symmetries of
the scalar sector of the SM --invariance under parity and charge conjugation--,
with its high energy interaction provided by the Lagrangian~\cite{Santos:2015,Pich-preparation},
%%%
%%%+\Delta\mL_{\rm non-res}^{(A)} $~\cite{Santos:2015,Pich-preparation}, with
\bear
\Delta\mL_V^{(A)} &=& \bra V_{\mu\nu}\,\, \mathbb{O}_V^{\mu\nu}\ket
\, ,
\qquad
%%%\mbox{being }
 \mathbb{O}_V^{\mu\nu} \,=\, \Frac{F_V}{2\sqrt{2}} f_+^{\mu\nu}
\,+\, \Frac{i G_V}{2\sqrt{2}} [u^\mu , u^\nu]
\,+\, \Frac{ c_1^V}{2}\,\left( \nabla^\mu J_V^\nu -\nabla^\nu J_V^\mu\right)/v^2
%%%+\, ...
  \, ,
\label{eq.example-L}
\eear
with $\bra...\ket$ for the matrix trace, $u_\mu= i u (D_\mu U)^\dagger u$, the combinations
$f_\pm^{\mu\nu} = u^\dagger \hat{W}^{\mu\nu} u \pm u \hat{B}^{\mu\nu} u^\dagger$
of the left and right field-strength tensors $\hat{W}^{\mu\nu}$ and $\hat{B}^{\mu\nu}$,
respectively, and $U=u^2=\exp\{ i \varphi^a\sigma^a/v\}$~\cite{Rosell:2012,Pich:2012dv}.
The precise definition of the covariant derivatives $D_\mu$
and $\nabla_\mu$ can be found in~\cite{Rosell:2012,Pich:2012dv}. The tensor
$J_V^\mu = - {\rm Tr_D}\{ \xi \bar{\xi} \gamma^\mu\}$
introduces the fermionic vector current
in a covariant way, with $\xi=u \psi_R+ u^\dagger \psi_L$
given by the $SU(2)_{R,L}$ doublets $\psi_{R,L}=\frac{1}{2}(1\pm \gamma_5)\psi$,
with $\psi=(t,b)^T$ (other SM doublets can be also added~\cite{Guo:2015})
%%%,flavor-ECLh})
and the Dirac trace ${\rm Tr_D}$.
The superscript $(A)$ refers to the
antisymmetric tensor formulation employed for the spin--1 resonance~\cite{Ecker:1988te}.
%We will later discuss the derivation in the 4-vector Proca formalism~\cite{Pich-preparation}.
%
%%%
%
The full Lagrangian may contain additional operators
not relevant for the form-factors analyzed in this talk~\cite{Pich-preparation}.
Integrating out $V$ one gets a contribution
to the EWET,  which at lowest order is given by
\vspace*{-0.5cm}
\bear
\label{eq.EWET}
\\
\Delta\mL_{\rm EWET}^{\rm from\, V} &=&
 \Frac{ \bra\mathbb{O}_V^{\mu\nu}\ket^2  }{2M_V^2}
 - \Frac{\bra \mathbb{O}_V^{\mu\nu} \mathbb{O}_{V\,\mu\nu}\ket   }{M_V^2}
%%%\left( \Frac{F_V}{2\sqrt{2}} f_+^{\mu\nu}
%%%\,+\, \Frac{i G_V}{2\sqrt{2}} [u^\mu , u^\nu]
%%%\,+\, c_1^V\, \nabla^\mu J_V^\nu /v^2\,
%%%+\, ...
%%%\right)
%%%\nn\\
%%%&=&
%%% \,
= \underbrace{   -\,i\, \Frac{F_V G_V}{4 M_V^2} }_{=\, i\, \mF_3/2}
\, \bra f_+^{\mu\nu} [u_\mu , u_\nu]\ket
\,\,\,
\underbrace{\, -\,
 \Frac{F_V c_1^V}{\sqrt{2} M_V^2} }_{=\mF^{X\psi^2}%_{10}
 }
\,  \bra f_+^{\mu\nu}   \nabla_\mu J_{V\,\, \nu}/v^2  \ket  \,\,
+\,\,  ...
\nn
\eear
with the dots standing for other effective operators not relevant in these proceedings.
For the Higgsless part, one has $\mF_3=a_2-a_3$
in Longhitano's notation of~\cite{Longhitano:1980iz,Morales:94}. In what follows, we will focus
on the Higgsless sector and $\mF_3,\mF^{X\psi^2},F_V,G_V$ and $c_1^V$
simply represent coupling constants.

The resonance Lagrangian~(\ref{eq.example-L}) provides the vector form-factors
of the $L+R$ current into two-Goldstones and
into two-fermions~\cite{Pich-preparation,Barbieri:2008,Rosell:2012,Pich:2012dv}:
\bear
\mathbb{F}^v_{\varphi\varphi}(q^2)\,=\, 1\,+\, \Frac{F_V G_V}{v^2}\, \Frac{q^2}{M_V^2-q^2}\, ,
\qquad\qquad
\mathbb{F}^v_{f\bar{f}}(q^2)\,=\, 1\,-\,  \Frac{ \sqrt{2} F_V c_1^V }{v^2} \, \Frac{q^2}{M_V^2-q^2}\, ,
\label{eq.VFFs-A}
\eear
with momentum transfer $q^\mu$.
The square form-factors $|\mathbb{F}^v_{ii}(s)|^2$
contribute to the $S$-parameter at one-loop
through the Peskin-Takeuchi sum-rule on the left-right correlator
$\Pi_{W^3B}$~\cite{Peskin:92}. If one requires that these form-factors give a ultraviolet-convergent
contribution
to the sum-rule, they must vanish at $q^2\to\infty$ and one obtains short-distance (SD)
constraints~\cite{Ecker:1989yg,Barbieri:2008,Rosell:2012,Pich:2012dv}
and predictions for the~LECs~\cite{Santos:2015,Pich-preparation,Ecker:1989yg}:
\bear
F_V G_V\, =\, v^2 \quad&\longrightarrow&\quad
\mF_3
=(a_2-a_3)
\,=\, -\, \Frac{F_V G_V}{2 M_V^2}
\quad \stackrel{{\rm SD \ constr.}}{=} \quad -\, \Frac{v^2}{2 M_V^2}
%%%\quad > \quad 13\cdot 10^{-3}
\, .
%%%\nn\\
%%%-\, \sqrt{2} F_V c_1^V\, =\, v^2 \quad&\longrightarrow&\quad
%%%\mF_{10}^{\psi^2 h^0}\,=\, -\, \Frac{F_V c_1^V}{\sqrt{2} M_V^2}
%%%\quad \stackrel{{\rm SD \ constr.}}{=} \quad  \Frac{v^2}{2 M_V^2}
%%%%%%\quad < \quad 13\cdot 10^{-3}
%%%\, ,
\label{eq.SD}
\eear
%%%where the function $F_V$, $G_V$ and $\mF_3$ of the Higgs field $h$
%%%are implicitly assumed to be evaluated at $h=0$ here and in the following~\cite{Santos:2015}.
For $M_V>1.5$~TeV    %%%~\cite{Pich:2012dv}
one finds the bound
\bear
-\, 1.3\cdot 10^{-2}\, \, < \,\,\,  \mF_3= (a_2-a_3) \, \, \, <\, \, 0\, \,  .
\eear
One can obtain analogous bounds for the LEC $\mF^{X\psi^2} =v^2/(2 M_V^2)$ by demanding
a similar SD behaviour
$\mathbb{F}^v_{f\bar{f}}(q^2)\stackrel{q^2\to \infty}{\longrightarrow} 0$
to the fermion form-factor, which would give $\sqrt{2} F_V c_1^V\, =\, -\, v^2$.

\subsection{$\mathbb{F}^v_{\varphi\varphi}$ form-factor: S-parameter}

The impact of the bosonic form-factor ${\rm F}^v_{\varphi\varphi}$
on the oblique parameters $S$ and $T$
was studied in a dispersive one-loop resonance analysis~\cite{Barbieri:2008,Rosell:2012,Pich:2012dv},
where the lightest triplet vector ($V$) and axial-vector ($A$) resonances were taken into account.
Therein, the contribution from the Goldstone and Higgs absorptive channels was incorporated.
In particular the ${\rm F}^v_{\varphi\varphi}(q^2)$ determined the contribution from
the $\varphi\varphi$ and $B\varphi$ cuts to the $S$ and $T$ parameter,
respectively~\cite{Pich:2012dv}.
%
%We considered the SD-constraints in~(\ref{eq.SD}) in addition to Weinberg sum-rules (WSR)
%for the $LR$ correlator $\Pi_{W^3B}$~\cite{WSR,Peskin:92}.
%
We studied asymptotically-free strongly coupled theories,
where $\Pi_{W^3B}$ satisfies the two Weinberg Sum Rules (WSRs),
and scenarios with weaker ultraviolet (UV) conditions (only the 1st WSR applies)
such as Conformal~\cite{Orgogozo:2012}  or Walking~\cite{WTC} Technicolour,
obtaining the 68\% confidence level determinations~\cite{Pich:2012dv}:
\bear
0.97\, <\, \kappa_W=M_V^2/M_A^2\,  <1\, ,&  \quad
M_V&\, >\, 5\, \mbox{TeV}
\quad (\mbox{1st \& 2nd WSR})\, ,
\\
 0.84\, <\, \kappa_W\,   <1.30\, ,&   \quad
M_V&\, >\, 1.5\, \mbox{TeV}
\,\,\,  (\mbox{only 1st WSR, for } 0.5<M_V/M_A<1  )
\, ,
\nn
\eear
where $\kappa_W$ denotes the $hWW$ (and $h\varphi\varphi$) coupling in SM units ($\kappa_W^{\rm SM} = 1$).

\subsection{$\mathbb{F}^v_{f\bar{f}}$ form-factor: $Z\to f\bar{f}$ anomalous couplings}

%The vector ($v_f$) and axial-vector ($a_f$) coefficients that parametrize
%
The $v_f$ and $a_f$ constants that parametrize
the $Z\to f\bar{f}$ decay have the form~\cite{Pich:2012sx},
\bear
v_f \,=\, T_3^f\, -\, 2 \, Q_f\,\sin^2\theta_W\,
+ \, (\delta g_R^{Zf}+ \delta g_L^{Zf}) \, ,
\qquad \qquad
a_f \,=\, T_3^f
\, + \, (\delta g_R^{Zf} - \delta g_L^{Zf}) \, ,
\eear
with $T_3^t=+1/2$, $T_3^b=-1/2$, the electric charge $Q_f$,
the weak angle $\theta_W$ and the new physics parametrized through the $\delta g_{R,L}^{Zf}$,
given in our low-energy description by
%%%.
%%%The contribution from the vector exchanges to the anomalous couplings
%%%$\delta g_{R,L}^{Zf}$ are given in our case by the $\mF^{\psi^2 h^0}_{10}$
%%%effective coupling~\cite{Pich:2012dv}, giving the tree-level estimate
\bear
|\delta g_{R,L}^{Zf}|%%%\,\bigg|_{\rm from V}
\, =\,
 |\mF^{X\psi^2}|\, \cos(2\theta_W)\, m_Z^2/v^2
%%%
%%%\quad\stackrel{\rm SD-constr.}{=}\quad
%%%\Frac{\cos(2\theta_W)\, m_Z^2}{2 M_V^2}\quad \lsim \quad 10^{-3}
\, ,
\eear
in agreement with current bounds of $\cO(10^{-3})$~\cite{deltag-exp}
for the fermion coupling
$ \mF^{X\psi^2} \,\sim \,   v^2/(2 M_V^2)< 1.3\cdot 10^{-2}   $
that one gets from the previous resonance coupling estimate $\sqrt{2} F_V c_1^V=-v^2$,
the bound $M_V>1.5$~TeV~\cite{Pich:2012dv} and the experimental value
$\cos(2\theta_W) \, m_Z^2/v^2= 0.07$.
%
%This result applies both for $Z\to b\bar{b}$ and $Z\to t\bar{t}$, avoiding
%large corrections in both cases, as one  has the generic
%chiral suppression $\mF_k  p^2/v^2 \sim p^2/M_R^2$.
%%%
%%%In the last step we
%%%use the SD-constraint that stems from requiring
%%%${\rm F}^v_{f\bar{f}}\stackrel{q^2\to\infty}{\longrightarrow} 0$.
%
%If the interactions between the composite states and the elementary fermions
%have a stronger chiral suppression~\cite{Pich-preparation,Buchalla:2013rka}
%this LEC $\mF^{X\psi^2}$ and the anomalous $Z f\bar{f}$ couplings $\delta g_{R,L}^{Zf}$
%can be even smaller.

\section{Equivalent Proca four-vector representation}

Through an appropriate duality transformation in the generating functional
it is possible to rewrite the underlying
resonance Lagrangian $\mL^{(A)}$ in~(\ref{eq.example-L})
%%%in terms of the antisymmetric tensor $V^{\alpha \beta}$
as a Proca Lagrangian $\mL^{(P)}$ in terms of four-vector field $\hat{V}_\mu$
and its field strength tensor
$\hat{V}_{\mu\nu}=\nabla_\mu \hat{V}_\nu - \nabla_\nu \hat{V}_\mu$.
A similar procedure~\cite{Pich-preparation,Ecker:1989yg,Bijnens:1995}
can be applied to models where the resonances are introduced
as gauge fields~\cite{gauge-resonances}.   %%%Feruglio:1988}.  %%%~\cite{Contino,Feruglio:1988}.
In the process, additional non-resonant operators with only light dof
are generated, which guarantee a proper UV behaviour.~\cite{Ecker:1989yg,Barbieri:2008,Bijnens:1995}.
On-shell, this duality can be read as $V^{\alpha\beta}=\hat{V}^{\alpha\beta}/M_V$
and $\nabla_\rho V^{\rho\mu}= - M_V\hat{V}^\mu$.
In our particular case, the duality transformation~\cite{Pich-preparation,Bijnens:1995}
changes the antisymmetric tensor
Lagrangian~(\ref{eq.example-L}) into
\bear
%%%\Delta\mL_V^{(A)} +\Delta\mL_{\rm non-res}^{(A)}
\mL^{(A)} \longrightarrow
%%%\Delta\mL_V^{(P)} +\Delta\mL_{\rm non-res}^{(P)}
\mL^{(P)}
 =&&  \bra \hat{V}_{\mu\nu}\,\,\left( \Frac{f_{\hat{V}}}{2\sqrt{2}} f_+^{\mu\nu}
+\Frac{i g_{\hat{V}} }{2\sqrt{2}} [u^\mu , u^\nu]\right)
+ \hat{V}_{\mu}\,\,\left( \zeta_{\hat{V}}\,  J_V^\mu/v^2  \,
%%%+\, ...
\right)\ket
\nn\\
&&
 -\, \bra \left( \Frac{f_{\hat{V}}}{2\sqrt{2}} f_+^{\mu\nu}
\,+\, \Frac{i g_{\hat{V}} }{2\sqrt{2}} [u^\mu , u^\nu]\right)^2\ket
\, ,
\label{eq.example-L-Proca}
\eear
with $f_{\hat{V}}=F_V/M_V$, $g_{\hat{V}}=G_V/M_V$ and $\zeta_{\hat{V}}= c_1^V M_V$.
In the low-energy limit $p\ll M_V$, Eq.~(\ref{eq.example-L-Proca}) leads to the same EWET,
\bear
\mL_{\rm EWET}  %%%^{\rm from\, \hat{V}}
&=&
  -\,i\, \Frac{f_{\hat{V}} g_{\hat{V}}}{4}
\, \bra f_+^{\mu\nu} [u_\mu , u_\nu]\ket
\,\,\,
  -\,
 \Frac{f_{\hat{V}} \zeta_{\hat{V}}}{\sqrt{2} M_V^2}
\,  \bra f_+^{\mu\nu} \nabla_\mu J_{V\,\, \nu}/v^2  \ket  \,\,
+\,\,  ...
\eear
%where the first term on the right-hand side is not originated now by the
%vector resonance exchange between $f_{\hat{V}}$ and $g_{\hat{V}}$ vertices
%(whose first contribution at $p\ll M_V$ is $\cO(p^6)$) but from the
%non-resonant piece $\Delta\mL_{\rm non-res}^{\rm SD}$
%that arises from the duality transformation $V\to \hat{V}$~\cite{Bijnens:1995,Pich-preparation}.
%Nonetheless, as expected, the low-energy theory
%and the predictions for $\mF_3$ and $\mF^{\psi^2 h^0}_{10}$ are the same regardless
%the chosen formulation for the spin--1 resonances.
The same agreement is found for the two form-factors previously obtained in~(\ref{eq.VFFs-A}):
\bear
{\rm F}^v_{\varphi\varphi}(q^2)\,=\, 1 \, +\, \Frac{f_{\hat{V}} g_{\hat{V}} }{v^2} q^2
\,+\, \Frac{f_{\hat{V}} g_{\hat{V}} }{v^2} \Frac{q^4}{M_V^2-q^2}
%%%\, =\, 1 \, +\, \Frac{f_{\hat{V}} g_{\hat{V}} M_V^2}{v^2}  \Frac{q^2}{M_V^2-q^2}
\, ,
\qquad
{\rm F}^v_{f\bar{f}}(q^2)\,=\,
1 \, - \, \Frac{ \sqrt{2}  f_{\hat{V}} \zeta_{\hat{V}}}{v^2}  \Frac{q^2}{M_V^2-q^2}\, .
\eear
%%%which are identical to those obtained in~(\ref{eq.VFFs-A}) in the antisymmetric tensor formalism.

\section{Conclusions}

%In this talk we have studied the interaction between the SM fields
%(bosons $\chi$ and fermions $\psi$) and possible heavy composite resonances.
%The Lagrangian is made invariant under
%$SU(2)_L\times SU(2)_R$.
%We construct all the possible $\cO(p^2)$ operators that are linear in the resonance
%fields $R$~\cite{Santos:2015,Pich-preparation}.
%This provides the most general resonance contribution to the
%low-energy EFT at NLO, this is, at $\cO(p^4)$.
%
The EWET couplings can be predicted in terms
of  resonance parameters; different resonance quantum numbers lead to different
patterns for the LECs~\cite{Santos:2015,Ecker:1988te,pseudovector-Cata}.
Further assumptions about the UV structure of the underlying theory
can be used to refine the predictions~\cite{Santos:2015,Pich:2012dv}.
In this talk we have provided a couple of examples
(oblique parameters $S$ and $T$and the anomalous $Zf\bar{f}$ couplings) to show
that composite resonances with masses of a few TeV ($M_R\sim 4\pi v\approx 3$~TeV)
are compatible with present direct and indirect searches.
%
%can easily comply with the low-energy electroweak precision tests thanks to
%
The $SU(2)_L\times SU(2)_R$ chiral invariance of the ECLh leads to
an appropriate low-energy suppression of tree-level NLO corrections
by factors $a_k p^2/v^2\sim p^2/M_R^2$
with respect to the LO prediction, $\cO(p^2)$~\cite{Santos:2015,Ecker:1988te,Ecker:1989yg}.
Finally, we have shown the equivalence between the antisymmetric tensors $V^{\mu\nu}$
and Proca four-vectors $\hat{V}^\alpha$ representations for spin--1 fields~\cite{Ecker:1989yg,Bijnens:1995}.

\end{document}